# In-series Multimode Interference Sensors and Fabry-Perot Interferometers for Enhanced Wavelength Shift Resolving Capabilities

João G. M. de Carvalho [1], Luiz D. C. Silva [1], Flavio A. M. Marques [1], Alexandre A. C. Cotta [1], Jefferson E. Tsuchida [1], Julio C. Ugucioni [1], Silésia C. da Silva [1], Leomar S. Marques [1], Diego C. Fuzatto [1], Alexandre Bessa dos Santos [2], Cristiano M. B. Cordeiro [3], Limin Xiao [4], Jonas H. Osório [1]

[1] *Multiuser Laboratory of Optics and Photonics, LaMOF, Department of Physics, Federal University of Lavras, Lavras, 37200-900, Brazil*
[2] *Federal University of Juiz de Fora, Juiz de Fora, 36036-900, Brazil*
[3] *Institute of Physics Gleb Wataghin, University of Campinas, Campinas, 13083-859, Brazil*
[4] *Advanced Fiber Devices and Systems Group, Fudan University, Shanghai, 200433, China*

*Abstract* — We report on the development of a refractive index sensor obtained by using a singlemode-multimode-singlemode (SMS) structure and a Fabry-Perot interferometer (FPI) set into an in-series configuration. Due to the self-imaging phenomenon, the SMS structure – formed by splicing a no-core fiber between two singlemode fibers –, provides a broad spectral peak whose central wavelength position is sensitive to variations in the refractive index of the medium surrounding the fiber. In turn, thanks to the in-series SMS-FPI configuration, the sensor's reflection spectrum exhibits the SMS spectral signature modulated by FPI fringes. This readily allows for reducing the width of the spectral features monitored during the sensing measurements, thus enhancing the capabilities of adequately resolving the corresponding spectral shifts. The FPIs reported in this investigation have been fabricated by using two different methods, namely by forming an air-gap FPI between the cleaved ends of two singlemode optical fibers, and by casting a polymeric film onto a connectorized fiber end tip. In the first configuration, the distance between the two cleaved fiber ends could be varied to tune the FPI's free spectral range, hence allowing for tailoring the widths of the spectral oscillations to be monitored during the sensing measurements. Alternatively, the second configuration, while avoiding the use of motorized translation stages, provides a more versatile option for applications. Thus, we understand that our work expands the application of multimode interference and FPIs in sensing scenarios, providing new opportunities for probing physical and chemical parameters by exploring their combined response.

*Keywords* — fiber sensors, multimode interference, Fabry-Perot interferometer, SMS structure

## I. INTRODUCTION

Optical fiber sensors have emerged as a powerful alternative to conventional electronic sensors due to their distinct advantages, including immunity to electromagnetic interference, small size, and high sensitivity. These characteristics make them excellent platforms for a wide range of applications, from structural health monitoring and industrial process control to medical and environmental sensing. The basic principle of most typical fiber sensors relies on detecting changes in light properties, such as intensity, phase, polarization, or wavelength shifts, as a function of external measurands. In this context, various configurations, including interferometers [1], fiber gratings [2], and specialty-fiber-based sensors [3], have been developed to translate variations in external parameters (*e.g.*, physical or chemical) into quantifiable optical signals.

Among the different fiber sensing technologies, devices based on multimode interference (MMI) have gained significant attention due to their compact size, simple fabrication, and high sensitivity [4, 5]. A typical MMI structure is the singlemode-multimode-singlemode (SMS) configuration. This structure is fabricated by splicing a multimode fiber (MMF) between two standard singlemode fibers (SMF). Light launched from the input SMF excites multiple modes within the MMF section. These modes propagate and interfere, leading to a phenomenon known as self-imaging at specific propagation distances, which results in a transmission spectrum with broad spectral peaks. The spectral position of these peaks can be sensitive to external parameters, such as temperature and the refractive index of the surrounding medium if the chosen MMF has its guidance properties dependent on the fiber external medium characteristics (*e.g.*, no-core fiber (NCF) [6], tapered fibers [7], or exposed-core fibers [8]). However, the broad width of these spectral peaks reduces the sensor's ability to detect wavelength shifts accurately.

In turn, Fabry-Perot interferometers (FPIs) constitute another class of highly versatile optical fiber sensors [1]. An FPI is formed by two spatially separated reflecting surfaces that create an optical cavity. Generally, in FPIs, light is partially reflected and transmitted at each interface, and the interference between these multiple reflections generates a fringe pattern in the reflection spectrum. The spectral position and spacing of these fringes are sensitive to changes in the optical path length of the cavity, which can be modulated by a variety of external parameters. In this context, FPIs have been extensively used as platforms for the realization of sensing measurements of pressure, temperature, and displacement [1].

In this paper, we report on a new approach to enhance the sensing capabilities of self-image phenomenon-based multimode interference sensors by setting SMS structures and FPIs into an in-series configuration. This combined SMS-FPI structure associates the inherent sensitivity of NCF-based SMS structures to external refractive index variations with the sharper spectral signature of the FPI. The reflection response of the proposed configuration is characterized by the typical SMS spectral signature modulated by FPI fringes, which allows for monitoring narrower spectral features rather than broad peaks during sensing measurements. To our knowledge, there is no report in the literature on combining the response of self-image phenomenon-based SMS configurations and FPIs for this purpose. In this context, here we demonstrate the proposed in-series SMS-FPI sensing platform by employing two distinct FPIs, namely an air-gap FPI, where the width of the spectral oscillations can be tuned by adjusting the cavity length, and a compact polymer-film FPI, which provides more versatility to the proposed sensor. Thus, we understand that this work presents a promising approach for developing new optical fiber sensors with tailorable spectral signatures and widens the applications of multimode interference and FPIs in sensing scenarios.

## II. SMS STRUCTURE CHARACTERIZATION

We begin by presenting the characterization experiments regarding the sensor elements individually, here starting with the SMS configuration. The SMS structure consists of a MMF section spliced between two SMFs. When light from the input SMF is launched into the MMF, it excites several modes that propagate along its length. As these modes interfere, the optical field can be replicated, leading to a phenomenon known as self-imaging [9]. The self-imaging distance, $Z_{SI}$, can be calculated by Eq. 1, where $n_{core}$ is the refractive index of the MMF core, $\lambda_0$ is the wavelength, and $D_{core}$ is the diameter of the core mode in a high-contrast waveguide [9]. Thus, if the MMF section is set to a length $L_{MMF}$ equal to $Z_{SI}$, a transmission peak around $\lambda_0$ is expected in the configuration's transmission spectrum.

$$Z_{SI} = \frac{4 n_{core} D_{core}^2}{\lambda_0} \quad (1)$$

In this work, a NCF (*i.e.*, a simple silica rod), was used as the MMF section. In this configuration, the external environment hence acts as the fiber's cladding, making the NCF's guiding properties directly sensitive to changes in the refractive index of the surrounding medium [6]. The NCF used in this investigation has an external diameter of (80 ± 1) μm and a length of (25.0 ± 0.5) mm. The splices between the NCF and SMFs were performed using a Fitel S179 splicing machine in its manual splicing mode. To characterize the transmission properties of the SMS configuration, we employed the setup illustrated in Fig. 1a. An HBK BraggMeter optical interrogator was used to measure the transmission spectrum of the system. For the refractive index sensing experiments, we utilized a 3D-printed reservoir to adequately submerge the NCF into water-glycerol solutions of varying concentrations.

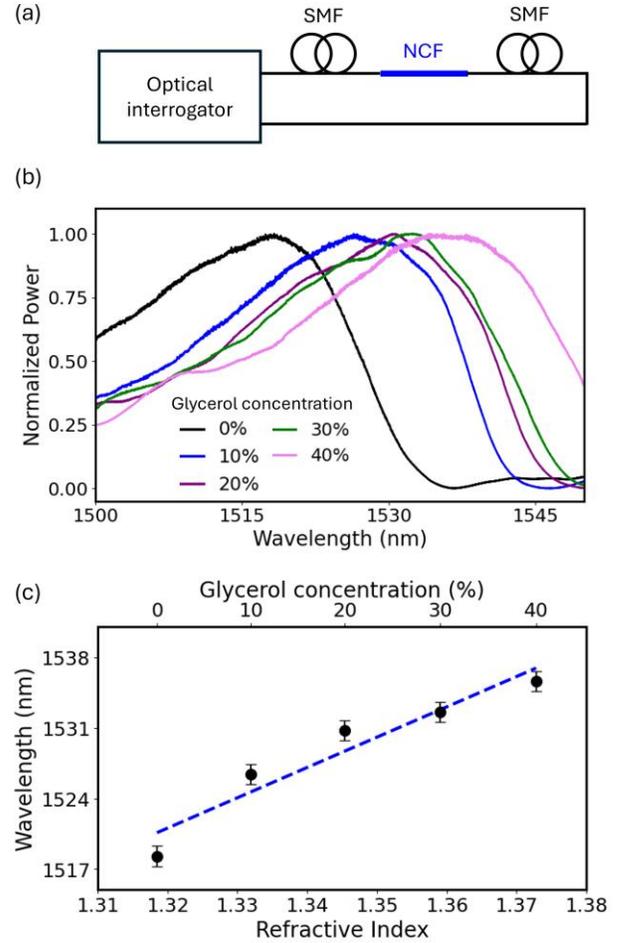

Fig. 1. (a) Diagram of the experimental setup used to characterize the SMS configuration transmission spectrum. (b) Measured transmission spectra of the SMS configuration for different external refractive indices, corresponding to varying glycerol concentrations in water-glycerol solutions. (c) The resulting wavelength redshift of the transmission peak as a function of the surrounding medium refractive index.

Fig. 1b shows the measured transmission spectra when the NCF was immersed in water-glycerol solutions with different concentrations (0% corresponding to distilled water), where one can observe the characteristic transmission passband of the SMS structure. Consistent with previously reported research [6], the transmission spectral peak is observed to redshift as the refractive index of the external medium increases due to higher glycerol concentrations. We mention that the spectra have been normalized by their corresponding maximum value for better visualization. In the spectra shown in Fig. 1b, the full width at half maximum (FWHM) of the SMS structure transmission peak has been accounted for as (31 ± 2) nm (the uncertainty in this measurement was determined from the standard deviation of the FWHM values corresponding to the peaks in the curves shown in Fig. 1b).

In turn, Fig. 1c exhibits the centroid wavelength of the transmission peaks as a function of the refractive index of the surrounding solution. From this data, a sensitivity of (300 ± 50) nm/RIU was determined. The refractive index values of the water-glycerol solutions, $n$, were calculated using Eq. (2), where $g$ is given by Eq. (3) [10], considering the refractive indices of water and glycerol at 1525 nm as given by their corresponding Sellmeier expressions [11, 12]. In Eq. (3), $n_g$ and $n_w$ are the refractive indices of glycerol and water. $C_g$ is the percentual volume concentration of glycerol.

$$n = \sqrt{\frac{2g+1}{1-g}} \quad (2)$$

$$g = \left(\frac{n_g^2 - 1}{n_g^2 + 2}\right)C_g + \left(\frac{n_w^2 - 1}{n_w^2 + 2}\right)(1 - C_g) \quad (3)$$

While the evaluation of the SMS structure's transmission spectrum allows for the realization of refractive index sensing, the corresponding capability of adequately resolving wavelength shifts is inherently limited, as the transmission peaks monitored during the measurements are considerably broad. In this context, the configuration proposed in this paper, which will be described in the following sections, addresses this limitation by integrating an SMS structure and an FPI in an in-series configuration. This arrangement allows for the reduction in the width of the spectral features, leading to an improvement in the capability of resolving the wavelength shifts during the sensing measurements.

## III. FPI CHARACTERIZATION

The second element in our sensing platform is a FPI. In this section, we thus present the characterization of the air-cavity FPI configuration used in one of the sensors to be described in the following. This FPI was obtained by cleaving the end-faces of two SMFs and by positioning them on the translation stages of a Fitel S179+ splicing machine. Using the machine's manual mode, we were able to gently approach the cleaved SMF tips until they touched. Subsequently, we created the FPI cavity by moving one of the translation stages back step-by-step (hence allowing for knowing the FPI cavity length, $L_{cav}$, according to the step displacements set in the splicing machine). For every translation step, the configuration's reflection spectrum was measured using the optical interrogator. Fig. 2a illustrates the experimental setup used for the FPI spectral characterization.

Fig. 2b shows representative reflection spectra measured for selected FPI cavity lengths. As expected by Eq. (4), we see in Fig. 2b that the free-spectral range (FSR) between consecutive fringes decreases as the cavity length is increased ($n_{air}$ is the refractive index of the air and $\lambda$ is the wavelength). In turn, Fig. 2c exhibits a graph of the measured FSR around 1525 nm as a function of $L_{cav}$, showing that the FSR is reduced from 24.7 nm to 6.9 nm when $L_{cav}$ is increased from 50 μm to 175 μm. The dashed line in Fig. 2c stands for the calculated FSR using Eq. (4), considering $\lambda$ = 1525 nm and $n_{air}$ = 1. Observation of Fig. 2c reveals good agreement between the measured results and the expected FSR values. We remark that the ability of FPIs to modulate the reflection spectrum of the system is fundamental to the proposed device's working principle. By placing the FPI interferometer in series with the SMS configuration, we can structure the SMS spectrum, enhancing the system's capabilities for detecting wavelength changes while maintaining its sensitivity to external refractive index variations.

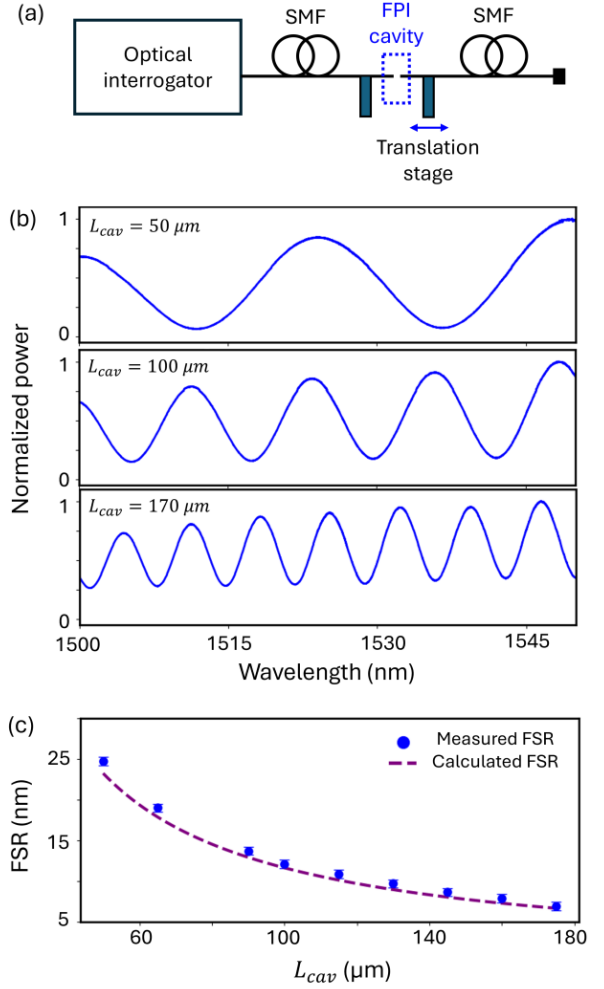

Fig. 2. (a) Diagram of the experimental setup used to characterize the FPI reflection spectrum. (b) Measured reflection spectra of the FPI for representative cavity lengths. (c) Free spectral range (FSR) as a function of the cavity length. The blue circles stand for the measured values, and the purple dashed lines represents the values calculated using Eq. (4), considering $\lambda$ = 1525 nm and $n_{air}$ = 1.

$$FSR = \frac{\lambda^2}{2\, n_{air}\, L_{cav}} \quad (4)$$

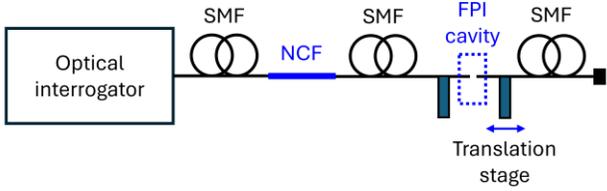

Fig. 3. Schematic diagram of the in-series SMS-FPI configuration. An optical interrogator launches a signal into the SMS section (SMF-NCF-SMF), which is then directed to an air-cavity Fabry-Perot Interferometer (FPI) with adjustable cavity length via a translation stage. The resulting modulated reflection spectrum is measured by the optical interrogator.

## IV. IN-SERIES SMS-FPI CONFIGURATION

The in-series SMS-FPI configuration we propose herein is schematically represented in Fig. 3. In this platform, the optical signal emitted by the interrogator is initially coupled to the SMS section, leading to a transmission spectrum as presented in Fig. 1 thanks to the self-imaging phenomenon (we clarify that, for the SMS-FPI demonstrations, we used the same SMS structure as characterized in Section I, but now measured in reflection). In sequence, the optical signal is directed to the air-cavity FPI, where it is reflected and guided back to the optical interrogator, where the reflection spectrum is measured. Such an in-series SMS-FPI structure hence leads to a reflection spectrum encompassing the characteristics of both platforms, namely a broad peak arising from the self-imaging phenomenon in the SMS configuration modulated by the interferometric fringes arising from the FPI. By changing the FPI cavity length (*i.e.*, by moving the cleaved fiber ends set on the translation stages of a splicing machine), the FSR of the fringes pattern can be altered, as we will show in the following.

Fig. 4 displays the measured normalized reflection spectra for the in-series SMS-FPI configuration corresponding to two representative FPI cavity lengths ($L_{cav}$ = 150 μm and 250 μm). Observation of the graphs allows for identifying that the characteristic broad spectral envelope generated by the self-imaging phenomenon in the SMS section is modulated by the interferometric fringes arising from the air-cavity FPI. The gray lines in Fig. 4 reproduce the corresponding transmission spectra of the SMS configuration, as presented in Fig. 1b, allowing for better visualization of the SMS-FPI modulated spectra (in Fig. 4, the transmission spectra of the SMS configuration have been used as a reference for normalizing the SMS-FPI spectra). As expected, as the FPI cavity length increases, the FSR of the spectral modulations decreases. This results in a higher density of fringes across the broad SMS peak, thereby structuring the spectrum and allowing for achieving enhanced capability of resolving the wavelength shifts during sensing measurements due to the decrease in the spectral oscillations FWHM.

Additionally, Fig. 4 exhibits the sensor's response to representative changes in the NCF's surrounding refractive index (corresponding to the situations in which the NCF has been immersed in three different glycerol-water solutions with distinct concentrations, 0%, 10 % and 20%). As the glycerol concentration increases from 0% (black curves) to 20% (purple curves), the envelope of the reflection spectrum is seen to redshift. Notably, as the cavity length increases and the fringes become sharper, the precision in identifying the corresponding wavelength shift (and thus the refractive index change) is improved.

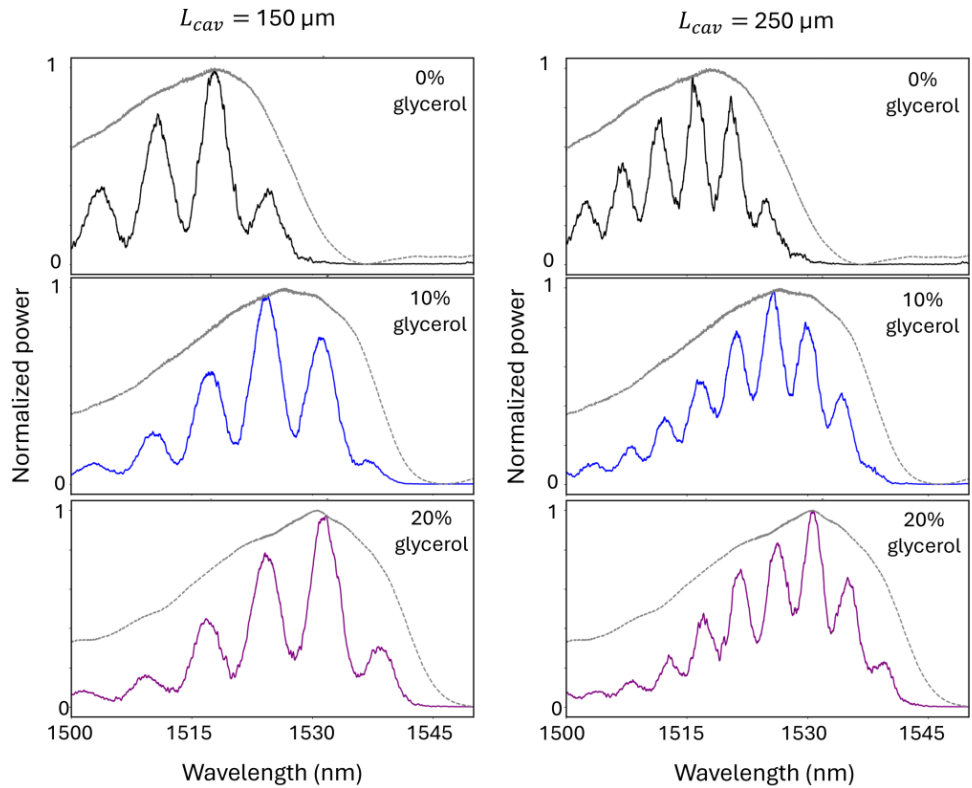

Fig. 4. Normalized reflection spectra for the in-series SMS-FPI configuration for representative cavity lengths ($L_{cav}$ = 150 μm and 250 μm) and different glycerol concentrations in the solution surrounding the NCF (black curves: 0% glycerol; blue curves: 10%; purple curves: 20% glycerol). The corresponding SMS transmission spectra (as presented in Fig. 1b) are shown as gray lines for comparison.

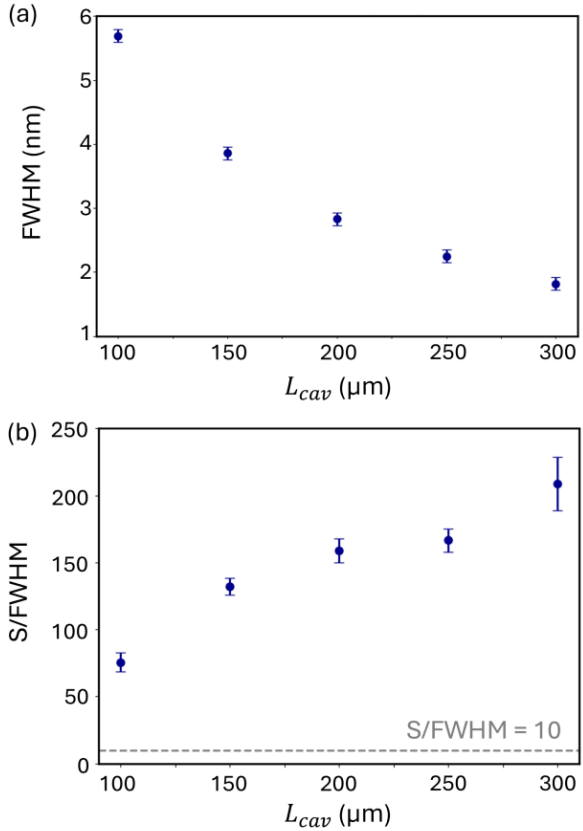

Fig. 5. (a) FWHM corresponding to the spectral fringes in the SMS-FPI reflection spectra as a function of the FPI cavity length ($L_{cav}$). (b) Ratio between the sensitivity and the FWHM of the spectral fringes (S/FWHM) as a function of $L_{cav}$. The horizontal dashed line refers to the S/FWHM estimated from the SMS transmission spectra (Fig. 2b).

In addition to the spectra shown in Fig. 4, we measured the response of the SMS-FPI configuration across a broader range of FPI cavity lengths. Fig. 5a exhibits the measured FWHM of the spectral fringes in the SMS-FPI reflection spectra as a function of $L_{cav}$. As anticipated from the reduction in the fringes FSR with increasing cavity length, the FWHM of the spectral fringes also decreases as $L_{cav}$ increases. Specifically, in our measurements, the fringes FWHM reduced from 5.7 nm at $L_{cav} = 100$ μm down to 1.8 nm at $L_{cav} = 300$ μm, which represents approximately a 5.4- to 17.2-fold reduction with respect to the FWHM of the SMS spectral peak when working in transmission, as characterized in Section I.

In turn, Fig. 5b plots the ratio between the measured sensitivity and the fringe FWHM (S/FWHM) as a function of $L_{cav}$. This ratio serves as a figure of merit, quantifying the system's overall performance by indicating an enhanced capability of resolving the spectral peak followed during the sensing measurements when S/FWHM is higher [13]. The results shown in Fig. 5b hence allows for identifying that, due to the FWHM reduction, one achieves greater S/FWHM values for larger $L_{cav}$. Remarkably, the S/FWHM values achieved in the in-series SMS-FPI configuration proposed herein allows for obtaining significantly higher S/FWHM values compared to that corresponding to the SMS configuration acting alone, here accounted for as $(10 \pm 2)$ and represented by a dashed horizontal line in Fig. 5b. At $L_{cav} = 300$ μm, for example, the S/FWHM corresponding to the SMS-FPI configuration reaches a value of $(210 \pm 20)$, which represents a 21-fold improvement compared with the simple SMS platform.

## V. IN-SERIES SMS-FPI CONFIGURATION WITH A POLYMER FILM AT THE FIBER END TIP

The configuration described in the last section allows for increasing the S/FWHM ratio compared to typical SMS structures by tuning the FPI cavity length. In this section, to make the system more convenient for practical applications by avoiding the need for translation stages for positioning the fiber tips and setting the FPI cavity, we describe an in-series SMS-FPI in which the FPI is obtained by casting a polymeric film on a connectorized fiber tip.

In the demonstration reported herein, to obtain the FPI cavity on the fiber tip, we employed an 18 wt% aqueous solution of poly(4-styrenesulfonic acid) (PSS-H) (Sigma-Aldrich, CAS 28210-41-5), selecting the material for its good film-forming properties. We remark that other materials could also be used to obtain the film-based FPI, as previously demonstrated in the literature [14, 15]. Here, the PSS-H film was obtained by drop-casting the PSS-H solution on a connectorized SMF (SC-PC fiber connector) and letting it dry overnight. Fig. 6a illustrates the procedure for the film preparation, and Fig. 6b shows a picture of the connectorized fiber with a PSS-H film on its tip. The experimental setup for characterizing the response of the in-series SMS-FPI is shown in Fig. 6c. Here, we maintain the same configuration as used previously (Fig. 3), with the only difference that, in this new demonstration, we changed the air-cavity FPI component to that made with the PSS-H film as described above.

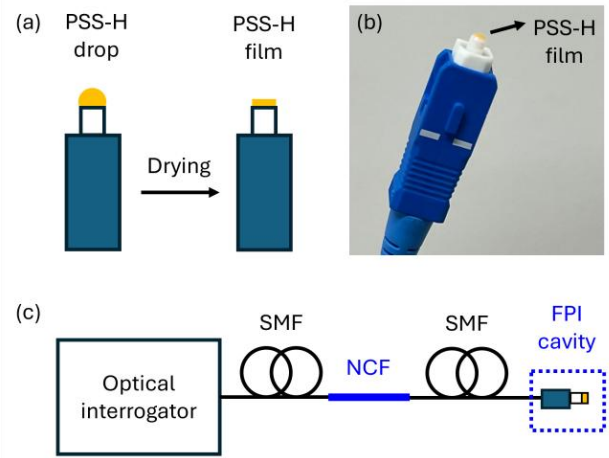

Fig. 6. (a) Illustration of the PSS-H film preparation on the connectorized optical fiber tip. (b) Picture of the connectorized fiber with the PSS-H film cast on its tip. (c) Representation of the experimental setup for characterizing the in-series SMS-FPI configuration.

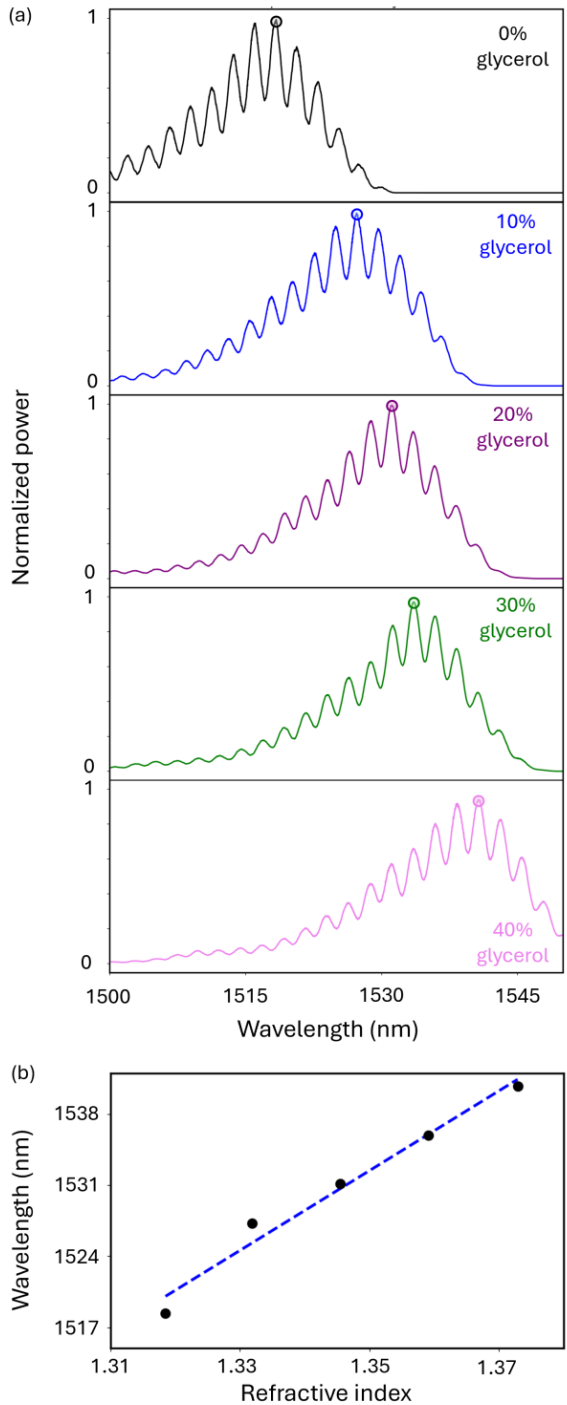

Fig. 7. (a) Reflection spectra of the in-series SMS-FPI configuration using the PSS-H film on the fiber tip as the NCF is immersed in glycerol-water solutions with different concentrations (the circles indicate the peak followed during the sensing measurements). (b) Wavelength redshift of the spectral maximum as a function of the surrounding medium refractive index.

To test the capability of the proposed configuration in sensing measurements, we immersed the NCF section in glycerol-water solutions with different concentrations while measuring the in-series SMS-FPI reflection response. The measured spectra are displayed in Fig. 7a, where one can recognize the SMS typical spectra modulated by the FPI fringes, and the corresponding spectral shift as the external refractive index is altered. Fig. 7b shows the wavelength redshift of the spectral maximum in the reflection spectra (as indicated by the circles in Fig. 7a) as a function of the external medium refractive index, from which we can calculate a sensitivity of $(400 \pm 30)$ nm/RIU. We observe that, in the sensing measurements reported here, we made the choice of following the wavelength shift corresponding to the spectral peak with maximum amplitude; however, it is worth mentioning that this analysis could also be performed by selecting different peaks.

In the spectra shown in Fig. 7a, we accounted for a typical FWMH of the spectral oscillations as $(1.2 \pm 0.1)$ nm. This hence entails an S/FWHM ratio of $(330 \pm 40)$. This represents a 33-fold increase in the S/FWHM metric compared with the SMS configuration acting alone. This result, combined with the tunability demonstrated in the air-gap FPI experiments, confirms that the in-series SMS-FPI architecture provides a versatile solution for enhancing the capabilities of resolving the spectral shifts in SMS-based optical sensors.

## VI. Conclusion

In this work, we proposed and characterized a refractive index sensor based on a SMS structure and a FPI arranged in an in-series configuration. Here, by exploiting the modulation of the SMS spectral signature with FPI fringes, we could successfully reduce the width of the monitored spectral features, thereby enhancing the capability to resolve wavelength shifts in refractive index sensing scenarios. We hence implemented this concept using two distinct FPI fabrication methods. First, we employed an air-gap FPI formed between cleaved fiber ends, which allowed for the tuning of the spectral oscillation widths by varying the cavity length. This configuration demonstrated that increasing the cavity length significantly reduces the FWHM of the fringes, yielding a 21-fold improvement in the sensitivity-to-FWHM ratio compared to the SMS structure alone. Second, we demonstrated a compact implementation of the sensor by casting a PSS-H polymer film onto a fiber tip. This approach resulted in an S/FWHM ratio 33 times higher than that corresponding to the basic SMS configuration, demonstrating the effectiveness of the approach proposed herein. These results thus confirm that combining multimode interference phenomena with Fabry-Perot interferometry is a promising strategy for developing high-performance optical fiber sensors with tailorable spectral characteristics for physical and chemical sensing.

## Acknowledgments

The authors thank the Minas Gerais Research Foundation, FAPEMIG (RED-00046-23, APQ-00197-24, APQ-01401-22, APQ-01618-25), the National Council for Scientific and Technological Development, CNPq (309989/2021-3,

305024/2023-0, 402723/2024-4, 409174/2024-6), Coordination of Superior Level Staff Improvement, CAPES, and the São Paulo State Research Foundation, FAPESP (2024/00998-6). The authors also thank E. S. Queiroz, D. L. Pereira, and S. T. S. Santos for their help with the PSS-H film preparation.